\documentclass[runningheads]{llncs}

\usepackage{footmisc}
\usepackage{graphicx}

\begin{document}
\title{Market Manipulation as a Security Problem}

\author{Vasilios Mavroudis}
\authorrunning{V. Mavroudis}
\institute{University College London\\
\email{v.mavroudis@cs.ucl.ac.uk}}

\maketitle

\begin{abstract}
Order matching systems form the backbone of modern equity exchanges,
used by millions of investors daily. Thus, their operation is strictly 
controlled through numerous regulatory directives to ensure that markets 
are fair and transparent. Despite these efforts, market manipulation 
remains an open problem.

In this work, we focus on a class of market manipulation techniques
that exploit technical details and glitches in the operation of the
exchanges (i.e., mechanical arbitrage). Such techniques are used by 
predatory traders with deep knowledge of the exchange's structure to 
gain an advantage over the other market participants. 
We argue that technical solutions to the problem of mechanical
arbitrage have the potential to significantly thwart these practices.
Our work provides the first overview of the threat landscape,
models fair markets and their security assumptions,
and discusses various mitigation measures.
\end{abstract}

\section{Introduction}
From outcry pits to electronic exchanges, trading has always been
a very competitive field, where even the slightest ``edge'' could be
used to gain an advantage over the rest of the market participants.
However, electronic trading gave rise to a previously unseen
type of arbitrage that exploits minor technical details and glitches in the
exchange's infrastructure~\cite{budish2015high}. 

These techniques are employed by traders with a very good understanding of the
exchange's operations (e.g., matching system's processes, order handling rules) and 
usually with low-latency access to the exchange's engine (on the order of milliseconds or less).
For example, traders have been found to exploit corner cases of the order-handling rules to unfairly
receive execution priority over other market participants~\cite{bodek2013problem}. Moreover, 
predatory traders use minuscule orders to uncover investors who submit hidden instructions
for large blocks of equity~\cite{budish2015high,budish2014implementation}.

While in retrospect this development was to be expected, the prevalence of such techniques 
caught markets off-guard, and led to various analyses and proposals by both researchers and 
regulatory bodies~\cite{angel2013fairness}. Our work complements these (legal and finance) 
research efforts and argues that these phenomena cannot be sufficiently studied without examining 
their technical component. 
We decouple the study of manipulation techniques from their legal standing, and model fair exchanges 
based on security properties and assumptions derived from EU and US market regulations.
On this basis, we examine six known mechanical manipulation techniques
and investigate how they violate one or more of the aforementioned properties.
The insights of this analysis are useful for the development of efficient countermeasures 
with minimum impact to the operation of the markets. 

To better understand the existing mitigation options and their effectiveness, we also discuss
and compare various technical and regulatory proposals. We conclude that technical countermeasures
cannot fully replace regulators but can significantly reduce the opportunities for manipulation.
This is also indicated by the small body of works introducing alternative market designs
that are less prone to certain forms of mechanical arbitrage~\cite{parkes2015achieving,CartlidgeSA18,budish2015high}. 

\medskip\noindent\textbf{Contributions.}
To summarize, this paper makes the following contributions:
\begin{itemize}
	\item We define the security properties that fair exchanges should satisfy and introduce the adversarial model assumed by the majority of modern equity exchanges.
	
	\item We study six known manipulation techniques, introduce their technical details, and investigate how they violate the exchange's security properties. To our knowledge, this is the first time that market manipulation techniques are studied in the ``systems security'' context.
	
	\item We survey existing countermeasures and discuss their effectiveness in tackling manipulation techniques based on mechanical arbitrage.
\end{itemize}

The rest of the paper is organized as follows: Initially, we detail the operation of order
matching systems and discuss the different order types (Section~\ref{sec:trading}). Subsequently, 
we list the basic security properties that a fair exchange should provide, and 
introduce a realistic model of an intelligent adversary (Section~\ref{sec:considerations}).
Section~\ref{sec:manipulation} studies the technical details of mechanical arbitrage techniques, while Section~\ref{sec:countermeasures} discusses both technical and regulatory countermeasures. Section~\ref{sec:conclusion}
concludes the paper.

\section{Background}\label{sec:trading}
\begin{figure}
	\centering
	\includegraphics[width=0.70\textwidth]{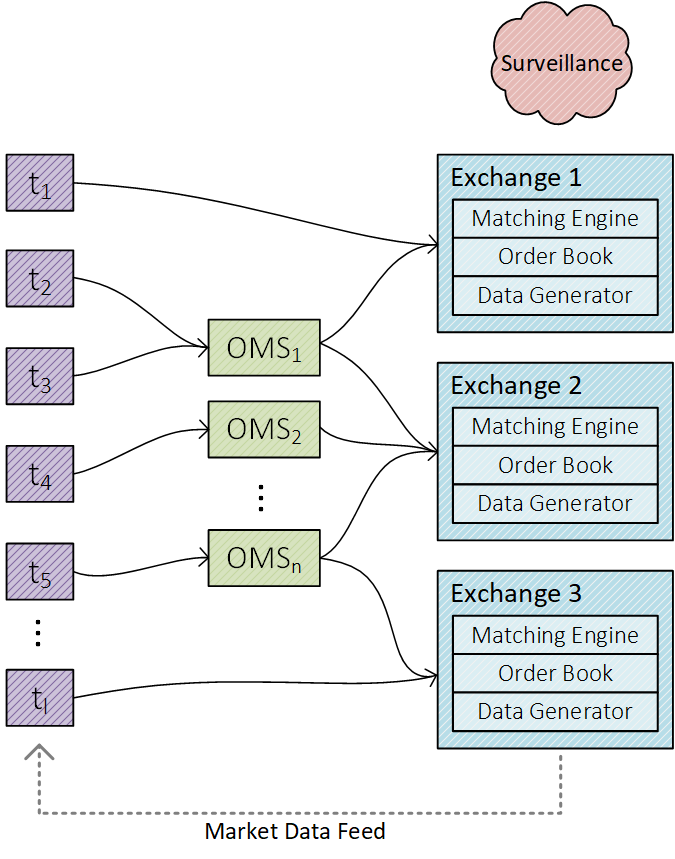}
	\caption{Illustration of the main actors participating in an electronic trading system. Traders ($t_i$) submit their instructions through the order management systems (OMSs) of their brokers who then forward them to the exchange with the best price. Alternatively, traders with direct market access can submit orders directly to the exchanges. In both cases, the matching engine of the exchange pairs buy and sell orders, and updates the order book with the remaining quantities. All these processes are monitored by the regulators to ensure the fairness and transparency of the markets.}
	\label{fig:structure}
\end{figure}

We now outline the basic subsystems of electronic exchanges and their operation.
Figure~\ref{fig:structure} illustrates the interactions between the different components and actors
of an electronic trading system. Initially, traders ($t_i$) submit their instructions (i.e., send, cancel, modify) 
to the order management system (OMS) of their broker. The OMS captures the details of each incoming instruction,
identifies the best execution venue, and routes it to that exchange\footnote{In practice, order routing is more complex, and optimal execution in a volatile market is non-trivial~\cite{foucault2008competition}.}.
The received orders are then placed in the exchange's order book, where the matching engine ranks, pairs and fills them with other sell or buy orders. An order is fully matched and cleared from the book 
if its entire open quantity is executed, while partially matched orders are updated to list only the remaining open quantity. Most equity and cryptocurrency exchanges operate continuous markets, where orders are matched on a continuous basis and the price is determined by the highest bid and the lowest ask quotes. 

Traders remain up to date with these prices either through the Stock Information Provider (SIP)\footnote{The Stock Information Provider is a central, consolidated stream and aggregator displaying the best priced bid and ask quotes, and the trading activity of each exchange\label{sip}.} or by subscribing to the trading data feeds offered by the exchanges. There are two types of such feeds: Layer 1 that provide only basic information such as real-time best bid/ask prices for securities trading in the exchange, and Layer 2 that offer quotes for all the orders resting in the order book (or up to a certain depth). In terms of latency, SIP averages at 0.09 milliseconds, while proprietary data feeds are several times faster~\cite{jones2018understanding}.

Finally, to ensure that traders, brokers and exchanges operate within the legal boundaries, regulators (e.g., the U.S. Securities and Exchange Commission) constantly surveil the market and investigate cases of abuse such as market manipulation and insider trading. Moreover, exchanges often monitor and review trader positions and transactions (i.e., self-regulate) to avoid facilitating illegal activities~\cite{IEXReg,NASDAQReg,NYSEReg}.

\medskip\noindent\textit{Dark pools}
are trading venues that protect the traders' privacy by not advertising the open orders in their book.
Pre-trade privacy is of great importance to institutional investors, as it enables them to submit large orders 
without adversely influencing the market price. Beside that, their operation is similar to that of standard exchanges (i.e., lit markets). Dark pools were initially used for equity trading, but the recent cryptocurrency burst gave rise to various new dark coin-trading venues~\cite{zhang2017republic}.

\subsection{Orders Types}\label{sec:ordertypes}
Trade orders are instructions sent by traders/investors to buy or sell on a venue,
and are either routed through a broker or sent directly to the exchange (when participants have direct market access). We now outline a few of the hundred order types (and variants) that modern exchanges support.

\medskip\noindent\emph{Market Orders.}
This order type specifies the quantity to be sold or bought, but not the price.
The exchange is responsible for filling the order at the best available price. 
While market orders are commonly used to quickly unload a ``position'',
traders prefer other types that give them better control over the execution price. 

\medskip\noindent\emph{Limit Orders.}
A limit order specifies a maximum or minimum price for buying or selling a number of shares. Unlike market orders, limit orders may not be executed if the price set cannot be met. Some variants allow investors to place time-limits for order execution after which the order is canceled.

\medskip\noindent\emph{Reserve Orders.}
Such orders are comprised of a displayed and a reserved (i.e., non-displayed) component (also called Iceberg orders). Once the displayed quantity is depleted to less than a round lot (i.e., the smallest order size allowed)
the reserved component is used to replenish it. There are several variants realizing different
replenishment strategies (e.g., fixed amount, random replenishment).

\medskip\noindent\emph{Discretionary Orders.}
This type of orders have both a displayed price and a non-displayed discretionary execution range.
The exchange first tries to fill the order at the displayed price, but if this is not
possible it looks for fills within the hidden discretionary range.

\medskip\noindent\emph{Anonymous Orders.}
These orders are used by investors who elect not to reveal their identity in a particular trade. 
Anonymous orders display a generic id in place of the unique participant's id.
Regulators and authorities monitoring the market retain access to the full details of the order.

\medskip\noindent\emph{Dark/Hidden Orders.} 
This class of orders provides varying degrees of pre-trade privacy, concealing parts of or all the details of an order before it is executed. They are available in both dark and lit markets and are used by investors wishing to buy or sell large volumes of stocks, options, or cryptocurrencies without influencing the market direction or sentiment. It should be noted that they do not provide any post-trade privacy, as this is prohibited by EU and US market regulations.

\subsection{Order Matching Algorithms}\label{sec:order}
Order matching is the process by which exchanges pair compatible buy and sell instructions.
For this purpose, each matching system uses an algorithm that considers several precedence criteria to determine which orders must be filled at any given time. This ranking process becomes of particular importance, when there are several open orders for the same security (or cryptocurrency) at the same price. Here, we outline the two most common categories of matching algorithms, while a more thorough analysis can be found in~\cite{janecek2007matching}.

\medskip\noindent\textbf{Price-Time Priority.}
Most equity and cryptocurrency markets use Time-based FIFO ranking.
FIFO is an intuitive ranking strategy, where orders are first prioritized based on their
offered price (i.e., a buyer that is willing to pay more is served first)
and then based on their relative submission time (in cases of orders with the same price).
For instance, given a buy order A for $200$ shares at a price $x$ that arrives before an order B for $50$ shares
at the same price, the system will seek to fully fill A first, and then move on to find matches for B.
Conditional order types further complicate the ranking process, as
ranking algorithms need to also consider several other order precedence rules.
For example, displayed orders (e.g., limit orders) have priority over all hidden order types (e.g., dark orders). Similarly, other ranking criteria are used to rank the different hidden order variants.

\medskip\noindent\textbf{Pro-Rata.}
As with price-time priority, Pro-Rata algorithms prioritize first the orders that offer the best price, 
but do not consider their relative submission time. Instead, they split the quantity sold 
between the buy orders at the same price level i.e., The volume of the shares allocated to each buy order is 
proportional to the number of available shares currently available at that price.
In our previous scenario, with two \textit{buy} orders A (for 200 shares) and B (50 shares) at the same price, an
incoming sell order for 200 shares will be matched with 160 shares from A and 40 shares from B (i.e., 80\% of each). In other words, if the offer does not suffice to cover the demand, all orders will be filled partially\cite{aspris2015time,guilbaud2015optimal}. Besides this basic version of the pro-rata algorithm, there are
several other variants with additional constraints such as maximum volume caps and minimum volume thresholds.

\section{Security Considerations}\label{sec:considerations}
We now examine market regulations to derive the security properties of modern exchanges and the threat model they operate under. We base our analysis primarily on literature for the US and EU markets, but our modeling remains relevant to exchanges worldwide.

\subsection{Security Properties}\label{sec:prop}
In 2005, the U.S. Securities and Exchange Commission (SEC) consolidated all pre-existing equity market regulations into the Regulation National Market System~\cite{RegNMS}, aiming to simplify market surveillance and increase investor protection.
In the European Union, the ``Markets in Financial Instruments Directive'' (MiFID II) provides a regulatory framework for investment services~\cite{mifid}. Based on these two regulations, we introduce a list of properties that, we believe, provide a well-rounded definition of fair and transparent markets. Comprehensive overviews of the U.S. Regulatory Landscape can be found in~\cite{parkes2015achieving,bodek2013problem}.

\begin{itemize}
	\item \textbf{Trading Integrity.}
	All the orders entered by market participants should express honest trading intent, and should not be used to artificially influence the volume of trades, the prices or any other market activity index.

	\item \textbf{Operational Transparency.}
	All the details and rules governing the operation of an exchange should be accessible by all market participants.
	
	\item \textbf{Fair Market Access.}
	All market participants should have equal access to the exchange when sending, modifying and canceling orders.

	\item \textbf{Symmetric Information Access.}
	All market participants should have equal access to up-to-date data on exchange's order book activity.
		
	\item \textbf{Order Queue Integrity.}
	Orders should be always ranked and executed according to the public order-matching rules of the exchange.
	
	\item \textbf{Participant Anonymity.}
	Non-authorized parties should not have access to the identity of traders using anonymous orders.
	
	\item \textbf{Data Confidentiality.}
	Non-authorized parties should not have access to the pre-trade data of dark orders. In the case of dark pools, other traders should not have access to any information related to a transaction (apart from the buyer and the seller).
\end{itemize}

The US and European market regulations assume that the majority of traders and exchange operators
abide by the regulations, but may occasionally violate one or more of the above properties even if it is explicitly prohibited by the law. Additionally, investors assume that other traders and exchange operators may operate in a legal but unethical manner (e.g., predatory trading).

\subsection{Adversarial Model}\label{sec:advmodel}
While in theory all traders have the same access to the market,
the complexity of modern exchanges provides various opportunities for
gaining competitive edges.
We now outline the main characteristics of intelligent traders that make
optimal use of their resources without overstepping any legal boundaries~\cite{budish2015high,aldridge2013high,menkveld2013high,hasbrouck2013low,kirilenko2017flash,pagnotta2018competing}:

\begin{itemize}
	\item \textit{High Computational Power.}
	They heavily and continuously invest in computer hardware and software, so as to process market data
	and execute orders as fast as possible. 
	
	\item \textit{Very Low Latency.}
	They minimize transmission latency by building their own high-bandwidth lines and by housing their computers in the exchange's premises, very close to the matching engine (i.e., colocation). All these can reduce transmission latency to under one millisecond~\cite{hasbrouck2013low,laughlin2014information}.
	
	\item \textit{Knowledgeable \& Strategic.}
	They have a very good understanding of the exchange's internal processes and systems, and make optimal 
	use of order types to leverage every (intended and unintended) feature they provide.
\end{itemize}

In finance literature~\cite{guilbaud2015optimal,adrian2016informational,clarke2014high},
high-frequency traders are often considered the most sophisticated class of market
participants, fitting all the aforementioned characteristics.

\section{Manipulation Techniques}\label{sec:manipulation}
We now describe several market manipulation techniques that exploit the underlying infrastructure of
exchanges and violate one or more of the properties listed in Section~\ref{sec:prop}. 

As part of a complex exploitation strategy, predatory traders 
often use attacks to: 1) deanonymize investors entering anonymous orders,
and 2) uncover large hidden orders.

\medskip\noindent\textbf{Latency Fingerprinting.}
This class of deanonymization techniques uses the order-transmission latency
as a side-channel to uncover the identity of investors that use anonymous orders (i.e., violates the Participant Anonymity property). The adversary maintains a \textit{latency table} that maps the different brokers/funds
to the time it takes for a message to reach the exchange from their network~\cite{adrian2016informational,sannikov2016dynamic}.
Upon observing an anonymous order and calculating its inter-arrival time, the adversary uses
the latency table to identify the broker that submitted it.
From a technical perspective, the transmission latency of an anonymous order is retrieved by 
subtracting the ``submission'' timestamp (found in each Financial Information eXchange 
Protocol message\footnote{\url{http://www.fixprotocol.org/}}) from the time the order was listed 
in the order book of the exchange. 

\medskip\noindent\textbf{Pinging.}
As outlined in Section~\ref{sec:ordertypes}, investors placing large orders often elect to 
use hidden order types to avoid influencing the market price and falling prey to predatory
practices. These orders are not displayed in the order book and are matched only if another 
order at the same price is submitted. Alternatively, investors often break up large buy orders 
into much smaller ones and enter them gradually into the market~\cite{aldridge2013high}. Pinging is one of the 
techniques used by predatory traders to uncover such investors, and is particularly effective in exchanges 
that send order execution notifications immediately and market activity updates at regular downsampled intervals.
The trader enters several sell orders of the smaller marketable size for the stocks they are interested in monitoring. Once one or more of those orders are matched, the trader is alerted (i.e., pinged) about the presence of a potentially large hidden buy-side order. This technique violates the Data Confidentiality property and
is usually combined with other techniques that exploit large buy orders (e.g., scalping).

\medskip\noindent\textbf{Quote stuffing.}
This technique is used only by sophisticated traders with direct access to the market (i.e., not through a broker), and involves placing and canceling high volumes of orders.
Its objective is to disrupt trading and create arbitrage opportunities by flooding the
data feeds of other participants, who are now unable to follow the market (Symmetric Information Access property)~\cite{egginton2016quote}.
Thus, it exploits bottlenecks in the data processing pipeline of regular investors.
In some cases, these attacks may also affect the responsiveness of the matching engine to market participants~\cite{mcnamara2016law,xie2016criminal}.

\medskip\noindent\textbf{Sniping.}
Sniping~\cite{budish2015high} is a form of mechanical arbitrage that relies on
high-speed transmission lines and co-location to gain an edge over slower investors.
Let a cryptocurrency or security traded on an exchange at price $x$ and a public signal \textit{y} that is strongly correlated to $x$. Consequently, when $y$ increases, traders immediately update their open 
sell quotes for that security from $x$ to a new increased price $\bar{x}$. However, this process is not instantaneous and hence momentarily (i.e., a few milliseconds) there are open sell quotes at the stale price $x$.
Predatory traders with high-speed connections exploit this delay, and submit buy
orders at that old price. With non-trivial probability~\cite{kirilenko2017flash}, some of their buy orders get filled at $x$, thus allowing them to immediately sell the security back for $\bar{x}$, risk-free~\cite{budish2015high}. This practice breaches the Fair Market Access property and has been also observed
in dark pools~\cite{adrian2016informational}.

\medskip\noindent\textbf{Scalping.}
This technique relies on very low-latency network links, violates Fair Market Access property, and is usually combined with market snooping techniques (e.g., pinging to identify large buy orders).
Let a broker handling a large buy order for 100,000 shares of company Z.
Upon examining the order books of all exchanges, the broker
finds 60,000 in exchange $E_1$ and 40,000 shares in exchange $E_2$, both at price x.
The broker submits a buy order for 100,000 shares to $E_1$ and gets a partial fill for 60,000 at x.
This also triggers a ``ping'' order set by a predatory trader, informing them that a large
order for Z is in the market (see also ``Pinging''). 
Following the order protection rule\footnote{The Order Protection Rule of Regulation NMS (Rule 611) introduced the National Best Bid and Offer requirement that protects investors from receiving sub-optimal prices for their quotes. In particular, Rule 611 mandates that exchanges should either reject marketable orders or route them to the exchange with the best price.}, $E_1$ now routes the order to exchange $E_2$ to buy the remaining 40,000 shares at price x.
However, the predatory trader utilizing its lower network latency, has already 
executed a buy order in $E_2$ for all available shares of Z at price x.
When the broker's buy order for 40,000 share arrives to $E_2$ (a few milliseconds after), the trader 
matches it with a sell order at a slightly inflated price $\bar{x}$~\cite{adrian2016informational}.

\medskip\noindent\textbf{Queue Jumping.}
While initially speed was sufficient for predatory traders to place their orders ahead of other market participants, the influx of technically advanced investors caused this practice to lose its effectiveness. 
To maintain their edge, predatory traders started using more sophisticated techniques that rely on special (i.e., ``exotic'') order types, allowing them to jump at the top of the queue\footnote{Orders that are closer to the top of the order book have access to more liquidity and better prices~\cite{cont2014price}.}, even if their order was not the first to enter the market.

In theory, these order types were accessible to all market participants and thus everyone could bump up the priority of their orders. However, in many cases exchange operators selectively disclosed them only to a subset of their clients and failed to inform everyone else about their existence and operation (Operational Transparency property)~\cite{ubsSEC,Buti13sub-pennyand,bodek2013problem,australiasic}. Nevertheless, 
these order types were exploitable only by sophisticated traders with knowledge of the precedence rules, very low-latency links and high processing power. 

\noindent\textit{Hide \& Light Orders}
is one of the most controversial order types. It was introduced as a workaround to Rule 610(d) of Regulation NMS~\cite{RegNMS}, which prohibits orders that lock\footnote{A buy order at price $x$ in an exchange $E_1$ locks the market, if there is another sell order for the same price $x$ in another exchange $E_2$~\cite{mcnamara2016law}.} or cross\footnote{A market is crossed if a buy order is posted at a price that is higher than the best (lowest) existing sell order, or inversely if a sell order is posted at a price lower than the best (highest) bid~\cite{mcnamara2016law}} the National Best Bid and Offer\footnote{The NBBO index is updated throughout the day with the lowest selling price and the highest buying bid (from all US exchanges) in order to ensure that investors receive the best possible price}.
To abide by this rule, exchanges temporarily adjust the price of the locking/crossing orders to be a tick (i.e., the minimum price movement) lower/higher than the NBBO, for as long as the market remains locked. Once the market unlocks, the prices are slid back to their original values. However, this does not apply to Hide \& Light orders, which retain their original price and are instead switched to being hidden (i.e., non-displayed). Once the market unlocks, they are automatically lighten up again. This seemingly subtle
difference corrupted the price-time priority of the queue, as lighten-up orders where placed in front of price-adjusted orders regardless of their relative ranking before the lock.

\medskip\noindent\textit{Day Intermarket Sweep Orders}
is another controversial order type, which is prioritized over all other types, including Hide \& Light orders\footnote{This prioritization rule concerns only the first Day ISO to enter the market and every subsequent Day ISO receives standard arrival time priority~\cite{bodek2013problem}.}. Day ISOs are based on a special exception in the Regulation NMS and are executed immediately without verifying if they lock or cross the market (the trader assumes all the regulation compliance liability). While their original motivation was to help institutional investors place large orders without chasing liquidity throughout the market, Day ISOs were quickly proven exploitable by predatory traders. In particular, ISOs were used by sophisticated traders with high-speed market activity feeds to rapidly react to market events, and place themselves ahead of regular investors that: 1) did not use Day ISOs, and 2) relied on the slow Stock Information Provider\footref{sip} to find the NBBO.

\section{Countermeasures}\label{sec:countermeasures}
Market regulation is an obvious measure of protection against predatory traders and manipulation practices. 
Until now, there have been several incidents where market operators and traders were accused of wrongdoing and settled to pay penalties~\cite{ubsSEC,australiasic,parkes2015achieving,CartlidgeSA18}.
However, keeping up with the latest manipulation practices, trading styles, market mechanisms, and distinguishing malicious-intent from irrational behavior is far from trivial~\cite{banks2010dark,soderstrom2011regulating}.
For this reason, a small body of work has proposed and, in some cases, deployed
technical countermeasures that aim to prevent all or some of these practices.
These countermeasures do not seek to replace regulators, but aim to alleviate
or at least reduce the opportunities for mechanical arbitrage.

As discussed in Section~\ref{sec:manipulation}, many of the manipulation techniques require 
that the trader can submit, alter and cancel orders within a few milliseconds.
For this reason, several exchanges introduced ``speed-bumps'' that delay the 
incoming orders and outgoing messages by a predefined amount of time. 
For example, the Investors Exchange (IEX) and the Toronto Stock Exchange (TSX)
use ``speed-bumps'' that introduce delay of 350 microseconds~\cite{buchanan2015physics}
and 1--3 milliseconds respectively~\cite{chakrabarty2018exchange,brown2015adverse}.
So far, ``speed-bumps'' have shown promising results in thwarting manipulation techniques
that rely on low latency. However, studies are not yet conclusive regarding their 
effects on the market~\cite{chakrabarty2018exchange}. Similarly,~\cite{budish2015high} introduces an alternative market design that uses discrete time instead of continuous. The authors argue that this design stops the arms race for speed and eliminates the risk-less latency-based arbitrage opportunities.
While most of the countermeasures require changes to the market design, 
there are certain steps that investors can take to protect themselves from
specific attacks. For example, ``latency tables'' (Section~\ref{sec:manipulation})
can be mitigated by adding noise on the timestamp of the submitted instructions.

Due to the above efforts, the rising infrastructure costs and the diminishing returns,
high frequency (i.e., very low-latency) trading activity has been steadily declining in 
the past few years~\cite{goldstein2014computerized}. Simultaneously, markets for other 
financial assets (e.g., currencies) have seen an increase in low-latency activity~
\cite{aggarwal2003stock,goldstein2014computerized}.

\section{Conclusions \& Future Work}\label{sec:conclusion}
This work aims to shed light on the technical aspects of market manipulation and to highlight the relevance of security analysis techniques and secure engineering principles to that problem. Toward this goal, 
we introduced a set of basic fairness properties, and used them to examine different 
market manipulation techniques and their mitigation counterparts. It is our hope that this new research direction will complement the legislative efforts for fair and transparent markets. While we focused on stock exchanges, future research could also investigate the risks that the average investor faces in unregulated cryptocurrency exchanges and decentralized trading platforms.

\bibliographystyle{splncs04}
\bibliography{bibliography}

\end{document}